\documentclass[useAMS,usegraphicx,usenatbib]{mn2e}
\usepackage{multirow}
\usepackage{rotating}
\usepackage{lscape}
\usepackage{longtable}
\usepackage{txfonts}
\usepackage{color} 
\title[HST/WFPC2 Imaging of the Dwarf Satellites And XI and And XIII]{HST/WFPC2 Imaging of the 
Dwarf Satellites And XI and And XIII : HB Morphology and RR Lyraes.$^{1}$}

\author[S-C. Yang \& A. Sarajedini]{S-C. Yang$^{2}$\thanks{E-mail: sczoo@astro.ufl.edu} and 
A. Sarajedini$^{2}$ \\
$^{[1]}$ Based on observations taken with the NASA/ESA Hubble Space Telescope, obtained at the Space Telescope Science Telescope.\\
$^{[2]}$Department of Astronomy, University of Florida, Gainesville, FL, 32611-2055, USA}

\begin{document}


\pagerange{\pageref{firstpage}--\pageref{lastpage}} \pubyear{2002}

\maketitle

\label{firstpage}

\begin{abstract}
We present a study of the stellar populations in two faint M31 dwarf satellites, 
Andromeda XI and Andromeda XIII. Using archival images from the Wide Field Planetary 
Camera 2 (WFPC2) onboard the Hubble Space Telescope (HST), we characterize the horizontal 
branch (HB) morphologies and the RR Lyrae (RRL) populations of these two faint dwarf satellites. 
Our new template light curve fitting routine (RRFIT) has been used to detect and characterize 
RRL populations in both galaxies. The mean periods of RRab (RR0) stars in And XI and And XIII 
are $<P_{ab}>$=0.621 $\pm$ 0.026 (error1) $\pm$ 0.022 (error2), and $<P_{ab}>$=0.648 $\pm$ 
0.026 (error1) $\pm$ 0.022 (error2) respectively, where ``error1'' represents the standard error of the 
mean, while ``error2'' is based on our synthetic 
light curve simulations. The RRL populations in these galaxies show a lack of RRab stars with high 
amplitudes ($Amp(V) > 1.0 $ mag) and relatively short periods ($P_{ab}$ $\sim$ 0.5 days), yet 
their period -- V band amplitude (P-Amp(V)) relations track the relation defined by
the M31 field halo RRL populations at $\sim$ 11 kpc from the center of M31. The 
metallicities of the RRab stars are calculated via a relationship between [Fe/H], 
Log P$_{ab}$, and Amp(V). The resultant abundances ($[Fe/H]_{And XI}=-1.75$; 
$[Fe/H]_{And XIII}=-1.74$) 
are consistent with the values calculated from the RGB slope indicating that our measurements are 
not significantly affected by RRL evolutionary away from the zero age horizontal branch. 
The distance to each galaxy, based on the absolute V magnitudes of the RRab stars, is
$(m-M)_{0,V}$=24.33 $\pm$ 0.05 for And XI and $(m-M)_{0,V}$=24.62 $\pm$ 0.05 for And XIII. 
We discuss the origins of And XI and And XIII based on a comparative analysis of the 
luminosity-metallicity (L-M) relation of Local Group dwarf galaxies.
\end{abstract}


\section{Introduction}

Since the Sculptor and Fornax dwarf spheriodal (dSph) galaxies were first discovered by 
Shapley (1938), this class of dSph galaxies has drawn a significant amount of attention recently; this is because of their potential role in the process of galaxy formation and evolution as the basic building blocks of giant galaxies 
in the cold dark matter (CDM) paradigm with a cosmological constant ($\Lambda$, White \& Rees 1978; 
Hernquist \& Quinn 1988, 1989; White \& Frenk 1991).
The dSph galaxies are the least luminous and appear to be the most common type of dwarf galaxy.
It is generally believed that all dSph systems in the Local Group harbor ancient
stellar populations as old as the oldest Galactic Globular Clusters (GGCs) indicating that Local
Group galaxies may share a common epoch of early star formation (Grebel \& Gallagher 2004).
The total masses (dark matter + baryonic mass) of the Local Group dSph systems within their central 300 pc are generally on the order of $\sim$ $10^7 M_{\odot}$ (Mateo et al. 1993; Strigari et al. 2008; Walker et al. 2009)
,which is comparable to the masses of the most massive Galactic GCs (i.e. the most massive GGC, $\omega$ Cen, has a total mass of $\sim$ $10^6$ $M_{\odot}$).
Despite their narrow range of masses, these dSph galaxies exhibit a wide variety 
in their star formation histories, which was described by Mateo (1998) in his review as ``no two 
Local Group dwarfs have the same star formation history''.

With regard to the mass range of dwarf galaxies, the recent discovery of ultra-faint dwarf (uFd) 
satellites around the Milky Way (MW) and M31
has dramatically extended the lower mass limit of these systems
(Willman et al. 2005a,b; Zucker et al. 2006a,b; Belokurov et al. 2006, 2007; Sakamoto \& Hasegawa 
2006; Irwin et al 2007; Walsh, Jerjen, \& Willman 2007). These newly discovered uFd systems are 
generally fainter than $M_{V}$ =--8.0, significantly dark matter dominated (at least $M/L >$ 100), 
and are some of the most metal-poor stellar systems ($[Fe/H] < -2$) found in the Local Group 
(Simon \& Geha 2007). The presence of these uFd galaxies is predicted by cosmological simulations
that seek to model the formation of the first galaxies (Ricotti \& Gnedin 2005); these suggest that 
some of these faint galaxies may be ``fossils'' of the first galaxies in which the bulk of stars formed 
before the reionization of the Universe at z$\sim$7--10 (Bovill \& Ricotti 2009).

And XI and And XIII were first discovered by Martin et al. (2006, hereafter M06) along with And XII 
via a MegaCam survey with the Canada-France-Hawaii Telescope (CFHT). By summing up the flux 
of bright members in the upper part of the red giant branch (RGB), they calculated a lower limit for the 
absolute magnitude of these galaxies and obtained values in the range, 
--7.3 $< M_{V} <$ --6.4. A follow-up investigation based on Keck/DEIMOS spectra 
and Subaru/Suprime-Cam imaging was carried out by Collins et al. 
(2010, hereafter C10). Based on their data, C10 obtained a very low metallicity ([Fe/H]$\sim$--2) 
for both And XI and And XIII. C10 also attempted to measure the distance to these faint satellites 
using both the tip of the RGB and the horizontal branch (HB) magnitudes. Their results locate And XI 
at a distance of $\sim$760 kpc and And XIII at a distance in the range of 760--940 kpc.

What is particularly interesting about And XI and And XIII is that these galaxies appear to fill the 
gap between the canonical dSph and the newly discovered uFd populations in the 
Luminosity-Metallicity (L-M) relation of Local Group dwarf galaxies. 
In this paper, we present 
a comprehensive study of the stellar populations in these two faint M31 satellites using deep archival 
images taken with the Wide Field Planetary Camera 2 onboard the Hubble Space Telescope 
(WFPC2/HST). This paper is organized as follows. Section 2 provides a short description of the data 
set and photometry; Section 3 describes general trends in the color-magnitude diagrams (CMDs); Section 4 illustrates our RR Lyrae (RRL) detection method and the pulsation properties of the RRL populations found 
in each galaxy; Section 5 describes the metallicity measurements using two independent methods 
(RGB slope and RRL periods); Section 6 presents the distance measurement; Section 7 illustrate 
a detailed HB morphology analysis; while Section 8 discusses the L-M relations of our target 
galaxies and their possible origins.; lastly, Section 9 presents a summary of our results.

\section[]{Observations and Data Reduction}

The HST/WFPC2 images of And XI and And XIII are available in the HST archive 
(program ID : GO-11084). The central regions of each galaxy were imaged 16 times 
in the F606W ($\sim$V) filter and  22 (And XI) /  26 (And XIII) times in the F814W
($\sim$I) filter with an exposure time of 1200 s. The detailed observing log is 
summarized in Table 1. 

\begin{table*}
\centering
\begin{minipage}{140mm}
\caption{Observing Log. \label{tbl-1}}
\begin{tabular}{cccrllc}
\hline\hline
Object   & R.A. (J2000) & Dec (J2000)  & Filters & Exp Time           & Data Sets          & HJD Range (+2 454 000)  \\
\hline
And XI   & 00 46 20.0   & +33 47 30.0  & F606W   & 16$\times$ 1200s   & u9x701-u9x702      & 352.78677 - 353.26802   \\
         &              &              & F814W   & 22$\times$ 1200s   & u9x703-u9x705      & 350.12355 - 364.38978   \\
         &              &              &         &                    &                    &                         \\
\hline
And XIII & 00 51 50.87  & +33 00 18.20 & F606W   & 16$\times$ 1200s   & u9x711-u9x712      & 304.23383 - 305.44764   \\
         &              &              & F814W   & 26$\times$ 1200s   & u9x713-u9x715      & 303.76789 - 314.70330   \\
         &              &              &         &                    &                    &                         \\
\hline
\end{tabular}
\end{minipage}
\end{table*}

We photometered the point sources in the WFPC2 images with the HSTphot package 
(Dolphin 2000) using the identical procedure adopted by Yang \& Sarajedini (2010). 
Pre-constructed point spread functions (PSFs) for each WFPC2 passband 
in the TinyTim PSF library were used to facilitate the PSF photometry. Bad pixels, cosmic rays, 
and hot pixels were removed by using the utility software included in the HSTphot package 
before performing photometry on the images. Aperture corrections, which are defined by the average 
difference between the PSF photometry and aperture photometry with a 0.5 arcsec radius, 
were calculated using the default settings of HSTphot. The instrumental magnitudes were 
transformed to the ground- based JohnsonÐCousins VI system using equations included in
HSTphot. The final list of stars from the resultant photometry 
includes only the point sources flagged as good stars (Object type 1) by the HSTphot classification.

\begin{figure}
\includegraphics[width=8cm]{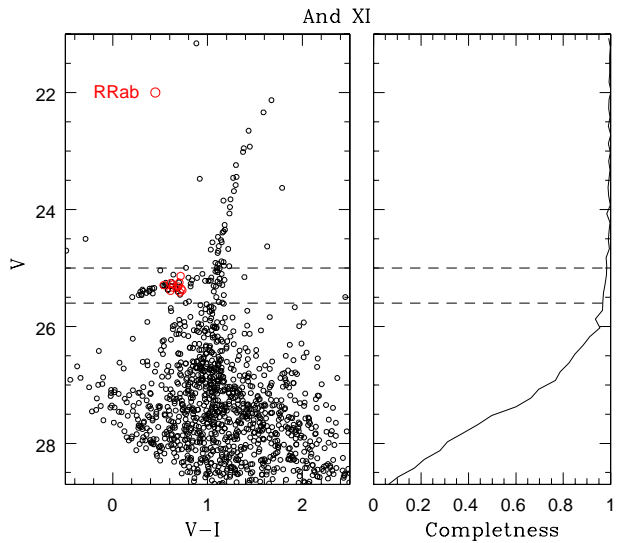}
\caption{The VI color-magnitude diagram of And XI. RRab candidates detected from our 
template light curve fitting routine (see section 4) are marked as red open circles. The right 
panel illustrates the photometric completeness indicating that our photometry is better
than $\sim$90\% complete at the level of the horizontal branch. \label{fig1}}
\end{figure} 

\begin{figure}
\includegraphics[width=8cm]{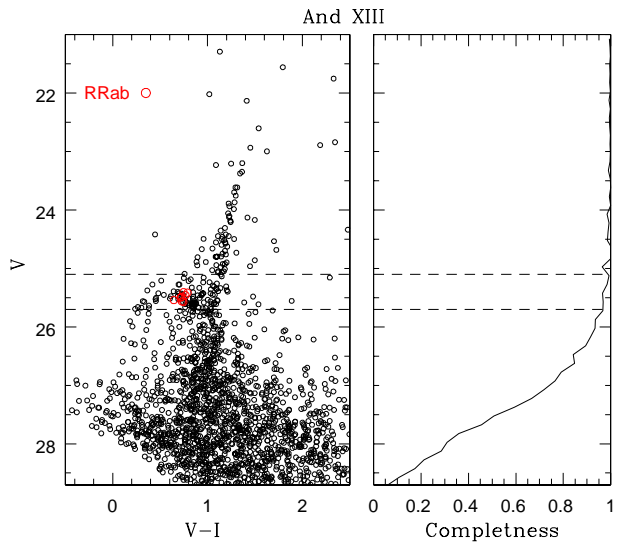}
\caption{Same as Figure 1, but for And XIII. \label{fig2}}
\end{figure}

\section{Color-Magnitude Diagrams}


Color-magnitude diagrams (CMDs) of And XI and And XIII in the VI passbands are presented in 
Figures 1 and 2. The right panels of each figure illustrate the photometric completeness derived from HSTphot's artificial star feature (Yang \& Sarajedini 2010), indicating that the point sources in both galaxies are well photometered to $\sim$1 magnitude below the HB level with better than $\sim$90\% photometric completeness. 
Thus, it is reasonable to assume that photometric incompleteness does not significantly affect the 
rest of our analysis, which is focused on the horizontal branch stars and brighter.

Both galaxies exhibit relatively steep and narrow red giant branches (RGBs) compared to the more 
luminous ($M_{V} < -8$) dwarf satellites in the Local Group. The HBs are well defined with several 
RR Lyrae candidates (see section 4).
The CMDs of both galaxies also exhibit no significant signs of young ($<$ 1Gyr) main sequence 
stars or intermediate age (1$\sim$10 Gyr) asymptotic giant branch populations. The overall features 
of the CMDs for And XI and And XIII are reminiscent of metal-poor Galactic GCs, indicating that these 
galaxies seem to be purely old and metal-poor stellar systems and have experienced relatively simple 
star formation histories (SFHs).

\section{RR Lyrae Variables}

\subsection{Detection and Characterization}
We searched for RRL stars in And XI and And XIII by employing a newly developed template 
light curve fitting method dubbed ``RRFIT (Robust RR Lyrae light curve FITing)'' written in 
FORTRAN. This new period searching routine works in a similar fashion to FITLC 
(Mancone \& Sarajedini 2008), which is our previous template light curve fitting code, but it 
is especially designed for large RRLs survey programs. Unlike other optimizing methods which use 
a gradient search or a brute force fitting routine to find the best-fit model parameters, 
RRFIT implements a robust genetic algorithm known as ``PIKAIA'', which efficiently finds the 
best-fit model parameters via an artificial intelligence random mesh searching technique. 
RRFIT takes full advantage of the PIKAIA algorithm, and as a result, significantly enhances 
computing speed (e.g. RRFIT is at least 10 times faster than FITLC) without sacrificing fitting 
accuracy. Another advantage of using RRFIT is the variety of standard templates that it can employ. 
In addition to the 8 RRL light curve templates employed by FITLC, we have adopted an 
additional 17 unique RRab templates from the work of Kovacs \& Kupi (2007) into our template 
light curve library to achieve a better description of various types of RRLs. 

In order to detect RRL populations in these faint dwarf satellites, we begin by identifying all of 
the variable star candidates within a range of V magnitude (24.5$ < V < $26.0) by calculating the 
reduced $\chi^2_{VI}$ value of each star defined by the following formula :  

\begin{displaymath}
\chi^2_{VI} = \frac{1}{N_V + N_I} \times
\Bigg[\sum_{i=1}^{N_V} \frac{(V_i - \overline V)^2}{\sigma_i^2} +  
\sum_{i=1}^{N_I} \frac{(I_i - \overline I)^2}{\sigma_i^2}\Bigg]
\end{displaymath}

Any peculiar data points that differ from the mean magnitude by $\pm$ 3 $\sigma$ were 
excluded from the $\chi^2_{VI}$ calculation. We compiled the list of potential variables by 
selecting stars with a $\chi^2_{VI}$ value greater than 3.0. For reference, the reduced 
$\chi^2_{VI}$ values of typical non-variable stars at the HB level (V(HB)$\sim$25.4) of 
And XI and And XIII is less than 3.0. This variability threshold generated a list of 69 
variable star candidates (46 in And XI; 23 in And XIII) from the WFPC2 
images of both galaxies. Then, we ran the RRFIT routine on the VI time series photometry 
of these variable star candidates in order to find the best-fit light curve parameters, which
include period, amplitude, and the epoch of maximum light. After careful eye-examination 
of the resultant light curves, we found 17 RRLs (10 RRab, 1 RRc and 6 unclassified) in 
And XI and 9 RRLs (8 RRab and 1 unclassified) in And XIII. The best fitting light curves 
for these RRL candidates from our RRFIT analysis are presented in Figure 3 and 4. The 
pulsation properties of the RRL candidates in each galaxy are listed in Table 2. The mean V 
band magnitudes of the RRab candidates from the best fitting light curves are 
$<V(RR)>$ = 25.31 $\pm$ 0.02 and 25.49 $\pm$ 0.02 for And XI and And XIII, respectively. 
The given uncertainties represent the standard error of the mean. We consider $<V(RR)>$ 
values as the best estimate of the $<V(HB)>$  value for each galaxy.  Several RRL candidates 
are noted as ``unclassified'' because the
goodness of fit yielded by the RRc and RRab templates are indistinguishable. These tend to
be the lower amplitude variables. A detailed discussion 
of these ``unclassified'' RRL candidates is presented in the following section.  
 
We also attempted to search for anomalous Cepheids (ACs) whose pulsation periods (0.4 $< P <$ 2 days) overlap significantly with the RRL period range. However, neither And XI nor And XIII appear to contain any ACs. Since ACs are produced by young-to-intermediate age (1--3 Gyr) populations, the lack of ACs in these galaxies supports our assessment of the overall star formation histories of these dwarf galaxies as inferred from their VI CMDs.

\begin{figure}
\includegraphics[width=8cm]{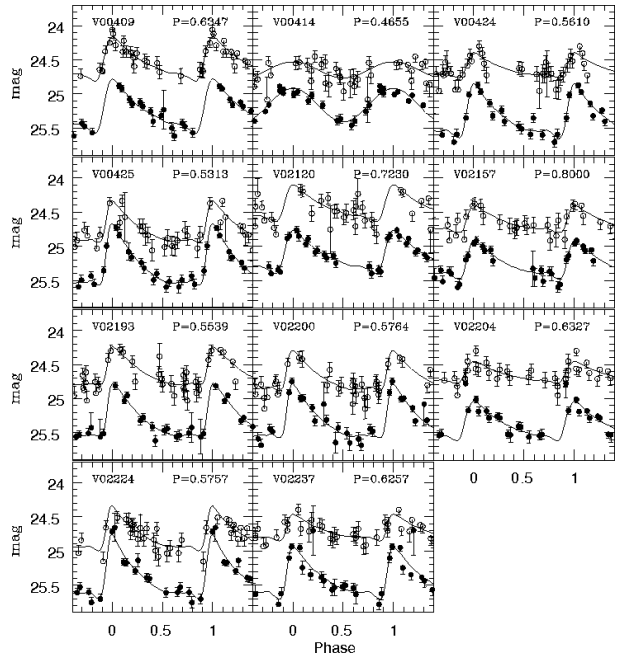}
\caption{The best-fit light curves of RRL candidates in And XI.\label{fig3}}
\end{figure}

\begin{figure}
\includegraphics[width=8cm]{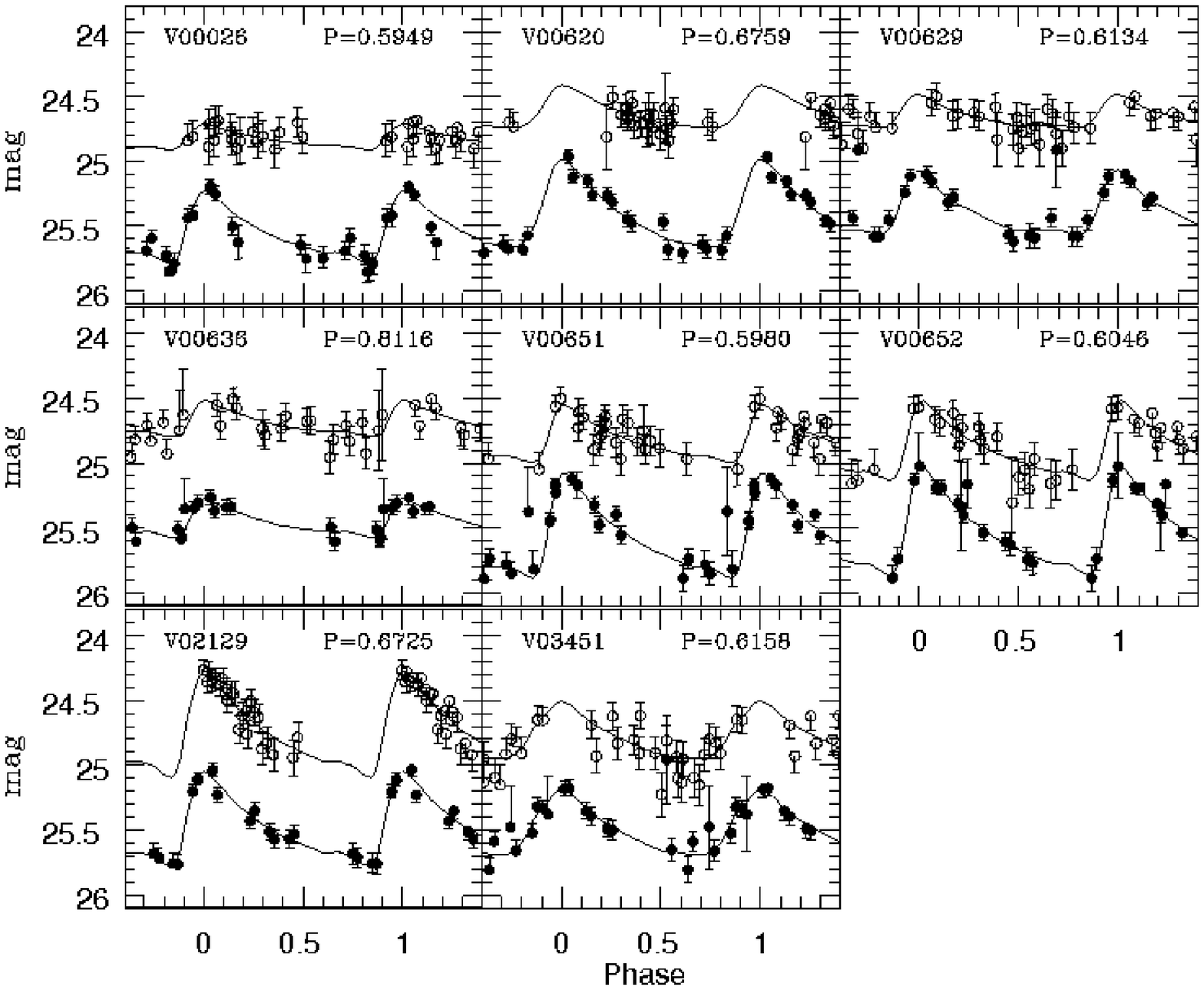}
\caption{Same as Figure 3, but for And XIII.\label{fig4}}
\end{figure}

\begin{table*}
\centering
\begin{minipage}{120mm}
\caption{Properties of RR Lyrae stars.\label{tbl-2}}
\begin{tabular}{lrrcccccc}
\hline
             &   R.A.       &     Decl.         &                   &
             &                  &    Period       &                   &                   \\
 ID        &  (J2000)  &   (J2000)      &                   &
  $<V>$ &  $<V>-<I>$ &  (days)         &

 A(V)     &  Type          \\

            &                     &                     &                &  And XI 
            &                     &                     &                &                     \\
\hline
V00409  & 0 46 16.47 & +33 46 59.86 &    & 25.2500 &  0.7032 & 0.6347 &  0.8044 &   ab  \\
V00414  & 0 46 16.13 & +33 47 19.47 &    & 25.1660 &  0.5168 & 0.4655 &  0.4905 &   c   \\
V00424  & 0 46 15.75 & +33 47 03.66 &    & 25.3583 &  0.7256 & 0.5610 &  0.7979 &   ab  \\
V00425  & 0 46 15.94 & +33 47 16.75 &    & 25.2951 &  0.5369 & 0.5313 &  0.9079 &   ab  \\
V00447  & 0 46 17.30 & +33 47 09.12 &    & 25.2636 &  0.5615 & 0.4290 &  0.6292 &   c?  \\
V02120  & 0 46 14.97 & +33 47 06.55 &    & 25.1377 &  0.7162 & 0.7230 &  0.6420 &   ab  \\
V02157  & 0 46 15.81 & +33 46 50.74 &    & 25.2600 &  0.6107 & 0.8000 &  0.5667 &   ab  \\
V02172  & 0 46 15.16 & +33 47 12.95 &    & 25.2149 &  0.6010 & 0.3852 &  0.3750 &   c?  \\
V02193  & 0 46 16.75 & +33 46 55.58 &    & 25.3169 &  0.6885 & 0.5539 &  0.7725 &   ab  \\
V02195  & 0 46 15.63 & +33 46 57.88 &    & 25.2808 &  0.6357 & 0.5080 &  0.4515 &   alias? \\
V02200  & 0 46 16.72 & +33 46 54.48 &    & 25.3353 &  0.6331 & 0.5764 &  0.7936 &   ab  \\
V02204  & 0 46 16.48 & +33 47 21.37 &    & 25.3839 &  0.7341 & 0.6327 &  0.5899 &   ab  \\
V02219  & 0 46 16.87 & +33 47 09.78 &    & 25.3640 &  0.5801 & 0.4275 &  0.6388 &   c?  \\
V02221  & 0 46 16.86 & +33 47 08.53 &    & 25.3588 &  0.6054 & 0.3793 &  0.5809 &   alias? \\
V02224  & 0 46 17.24 & +33 47 13.06 &    & 25.3926 &  0.6089 & 0.5757 &  0.9582 &   ab  \\
V02237  & 0 46 17.56 & +33 46 54.70 &    & 25.4184 &  0.7088 & 0.6257 &  0.7728 &   ab  \\
V04338  & 0 46 17.04 & +33 46 49.26 &    & 25.3653 &  0.6123 & 0.4070 &  0.5766 &   c?  \\
\\
\hline
\\
        &            &              &    & And XIII &        &        &         &       \\
\hline
V00026  & 0 51 46.00 & +33  0 04.38 &    & 25.5649 &  0.7299 & 0.5949 &  0.5677 &   ab  \\
V00620  & 0 51 46.18 & +33  0 11.12 &    & 25.4153 &  0.7894 & 0.6759 &  0.6772 &   ab  \\
V00629  & 0 51 44.58 & +33  0 08.32 &    & 25.4071 &  0.7464 & 0.6134 &  0.4953 &   ab  \\
V00636  & 0 51 46.99 & +33  0 08.71 &    & 25.4605 &  0.7675 & 0.8116 &  0.2972 &   ab  \\
V00651  & 0 51 46.81 & +33  0 07.11 &    & 25.5733 &  0.7460 & 0.5980 &  0.8240 &   ab  \\
V00652  & 0 51 46.55 & +33  0 16.74 &    & 25.5419 &  0.6450 & 0.6046 &  0.8455 &   ab  \\
V02129  & 0 51 46.35 & +33  0 13.60 &    & 25.4872 &  0.7252 & 0.6725 &  0.7452 &   ab  \\
V03451  & 0 51 46.88 & +33  0 09.63 &    & 25.4999 &  0.7173 & 0.6158 &  0.5103 &   ab  \\
V02147  & 0 51 46.48 & +33  0 12.96 &    & 25.3780 &  0.5491 & 0.4287 &  0.4970 &   c?  \\
\hline
\end{tabular}
\end{minipage}
\end{table*}

\subsection{Synthetic Light Curve Simulations}
In order to quantitatively gauge the influence of false periods (aliases) on our analysis, we 
have performed the following set of synthetic light curve simulations (Sarajedini et al 2009 (S09); 
Yang \& Sarajedini 2010; Yang et al 2010, hereafter Y10). Applying the same observing 
windows as the And XI and And XIII WFPC2 images, $\sim$ 1000 synthetic light curves of 
RRab and RRc stars were generated. Periods and amplitudes were randomly assigned to each 
artificial RRL within appropriate ranges for each type (RRab : $0.5 < P <1.2$ , $0.2 < Amp < 1.5$; 
RRc : $0.2 < P < 0.5$, $0.2 < Amp < 0.5$). Photometric errors for each data point were assigned 
using gaussian deviates centered at 0.06 mag, which is a typical photometric error at the level of the
HB for both galaxies in the WFPC2 photometry. Then, we ran the RRFIT routine on these artificial 
RRL stars. 

The plots shown in the top panels of Figure 5 and Figure 6 illustrate the difference between the input 
and output periods as a function of input period. For And XI (see Figure 5), RRc stars appear to 
experience more significant aliasing compared to RRab stars. Quantitatively, $\sim$ 84 \% of the input 
periods for RRab stars were accurately recovered within $\pm$ 0.05 day, while $\sim$ 61 \% of 
the RRc periods were recovered within the same $\pm$ 0.05 day period range. In addition, 
the output periods of the 
RRc stars tend to be longer than the input periods. The negative tail shown in the $\Delta$P distribution 
(the left panel in the middle of Fig 5) for the artificial RRc stars of And XI is mainly caused by this bias 
in the RRc periods. This result reveals that there can be considerable uncertainty in the properties of 
individual RRc stars in And XI. 

\begin{figure}
\includegraphics[width=8cm]{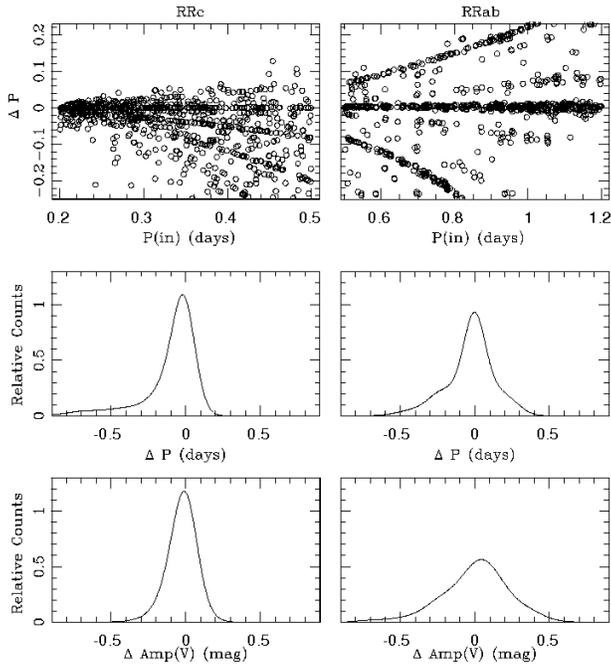}
\caption{The results of our synthetic light curve simulations for the And XI data. Top panels illustrate 
the difference ($\Delta$P=$P_{in}-P_{out}$) between the output period and the input period as 
a function of the input period for two different pulsation modes. 
Lower panels (middle \& bottom) show the distributions of the deviation of the output period 
and amplitude from the input values. Our simulations reveal that the resultant periods from our 
analysis are likely to have more aliasing in the RRc regime as compared to the RRab regime.  
\label{fig5}}
\end{figure}

\begin{figure}
\includegraphics[width=8cm]{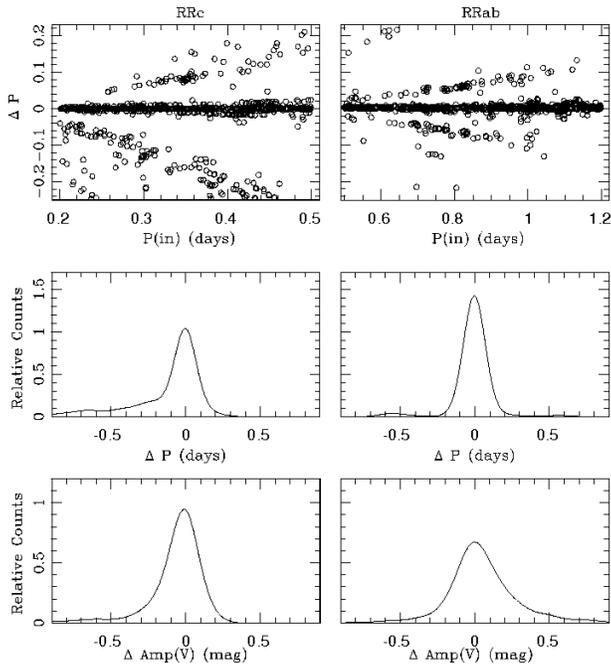}
\caption{Same as Figure 5, but for And XIII. The simulations show that the aliasing issue is 
less significant for the And XIII RRL candidates as compared with the And XI RRLs.\label{fig6}}
\end{figure}

To investigate the degree of misidentifications in the RRc regime, we performed a comparative 
analysis between the best-fit light curves and the first-overtone mode light curves for those 
unclassified RRL candidates found in the previous section. Observing cadence and phase 
coverage are the two most important factors for accurate period measurements (Y10), 
and these factors also have a significant role in determining which type of light curve provides 
the best fit to the observations.
Photometric errors also influence the appearance of the measured light curves. Large photometric 
errors can deform the original shape of a light curve and produce a misidentification of the type 
of variable (Yang \& Sarajedini 2010). However, based on our simulations, the 
measured periods and amplitudes are largely insensitive to the actual classification of the 
RRL (e.g. whether it is classified as an RRab or RRc). 
From Figure 7, we see that the majority of unclassified RRL stars exhibit 
good quality fits both in the fundamental and first-overtone modes. However, their
location in the Period-Amplitude (P-A) diagram (see Sec. 4.3) places them in the RRc regime. 
Hence, we conclude that these unclassified RRLs are likely to be RRc stars. 
There are two exceptions to this - V02195 \& V02221, which show significantly different 
solutions in the resultant periods
for each type of pulsation mode (see Figure 7). Again, relying on their location in the
P-A diagram, we suspect that the RRc template is more appropriate for these stars and that
the RRab period is the result of an alias. 
Therefore, the corrected census of RRc stars in each galaxy is $N_{c}$ = 5 and 1 for 
And XI and And XIII, respectively.

\begin{figure}
\includegraphics[width=8cm]{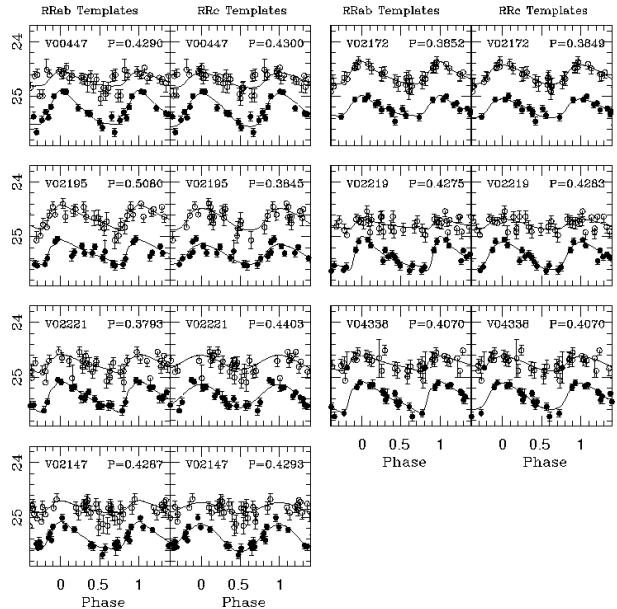}
\caption{Comparisons between the RRab and RRc template fits for the unclassified 
RRL candidates. Note these unclassified RRL tend to have smaller amplitudes than
most other RRLs in our sample. \label{fig7}}
\end{figure}

In order to estimate the errors of the individual RRL periods, we performed the following 
statistical test introduced in Y10. From the artificial RRL lists, we randomly draw the 
same number of artificial RRL stars as our observed RRLs in each galaxy 
[And XI : 10 (RRab) and 5 (RRc); And XIII : 8 (RRab) and 1 (RRc)] to calculate an 
average $\Delta$P value for each sample drawn. Note that since And XIII only has one
RRc star, we have excluded these And XIII stars from this analysis.
We construct the distribution of $<\Delta P>$ values for each pulsation 
mode by iterating this probability sampling 10,000 times.
Figure 8 illustrates the resultant $<\Delta P>$ distributions of the artificial RRL stars for 
And XI and And XIII 
along with the best-fit Gaussian distributions. The 1-$\sigma$ errors of the best-fit Gaussians, 
as given in each panel, are good approximations to the standard deviations of each $<\Delta P>$ 
distribution. Therefore, the standard errors ($\sigma/\sqrt{N}$) of the individual RRL periods in our 
analysis are $\sigma(P_{c})$=$\pm$0.038 and $\sigma(P_{ab})$=$\pm$0.022 days for the RRL 
candidates found in And XI. For the And XIII RRab candidates, we find 
$\sigma(P_{ab})$=$\pm$0.022 days. The same method 
was applied to estimate the errors in the amplitudes. We obtained 
$\sigma(Amp(V)_{c})$ = $\pm$ 0.027 and $\sigma(Amp(V)_{ab})$ = $\pm$ 0.024 mag 
for the And XI RRL candidates. For the And XIII RRab stars, the error value is
$\sigma(Amp(V)_{ab})$ = $\pm$ 0.033 mag. 
We found no significant biases in our determination of the mean periods and amplitudes. 
Hence, we assume that period aliasing does not significantly affect the results in the 
remainder of our analysis, which is focused on the mean periods and amplitudes.

\begin{figure}
\includegraphics[width=8cm]{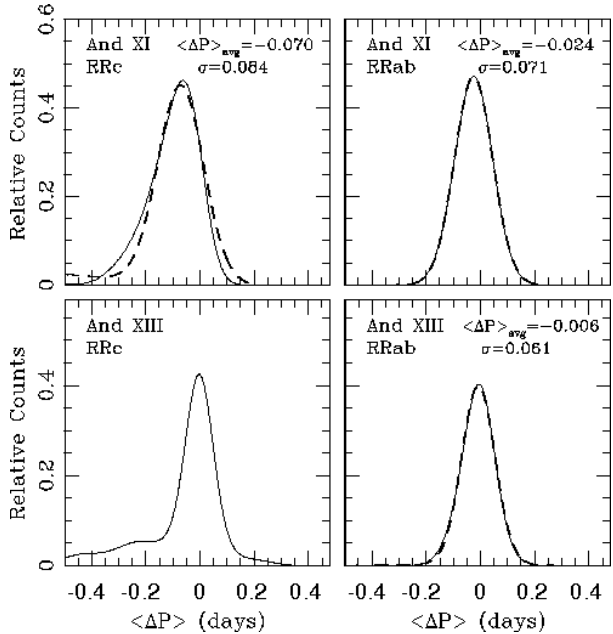}
\caption{The distributions of the output period deviation for synthetic RRab and RRc stars. 
The dashed lines illustrate the best-fit Gaussians for each distribution. $<P>_{avg}$ and $\sigma$ 
represent the peak and the standard deviation of the Gaussian fits, respectively.\label{fig8}}
\end{figure}

\subsection{Period-Amplitude (P-A) Diagrams}

The P-A relations of the RRL populations in And XI and And XIII are presented in Figure 9. 
For comparison, we generated contoured P-A diagrams for the RRL stars in the spheroid 
of M31 from S09 representing a population at $R_{gc}$ $\sim$ 5 kpc) and from
Brown et al. 2004 (B04, hereafter referred to as the Brown field) at 
$R_{gc}$ $\sim$11 kpc. The contour plot for the Brown field 
was scaled up by a factor of 5 in order to see its trend more clearly.  The RRL stars (open symbols) 
discovered in And XI and And XIII from our analysis are plotted on top of these contoured 
P-A diagrams. Two solid lines representing the Oosterhoff I (Oo I) and II (Oo II) sequences for 
the Galactic Globular Clusters (Clement 2000) are also plotted. Since the amplitudes of the 
M31 RRL stars in these two previous studies were determined in the ACS/F606W passband, 
we applied an 8\% increment to the amplitudes of each M31 RRL in order to convert the 
F606W amplitudes into V band amplitudes (B04; S09).

\begin{figure}
\includegraphics[width=8cm]{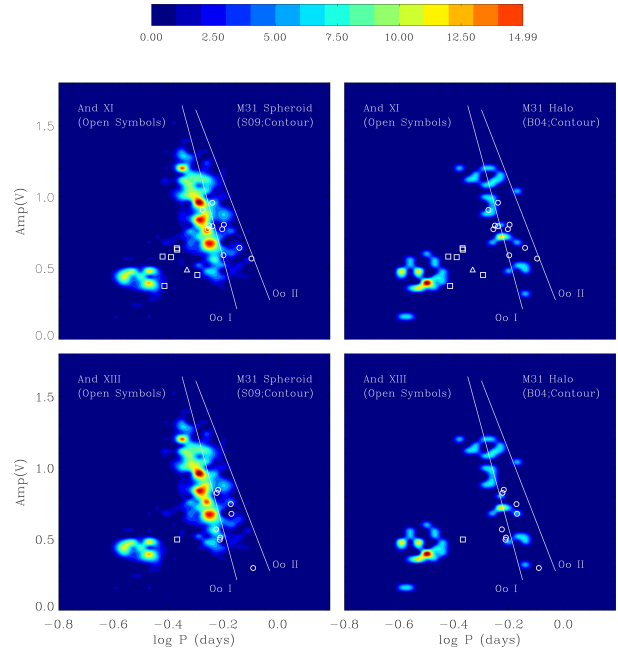}
\caption{The Period-Amplitude (P-A) relations of RRL candidates (open circles) in 
And XI and And XIII are plotted on top of the contoured P-A relations of the M31 
halo RRLs (left panels, S09 : inner halo, $R_{gc}$ $\sim$ 5kpc ; right panels, B04 : 
Brown field, $R_{gc}$ $\sim$ 11kpc). The contour plot for the Brown field 
was scaled up by a factor of 5 in order to see its trend more clearly.
Solid lines represent the fiducials of Oosterhoff I and II GGC (Clement 2000).
\label{fig9}}
\end{figure}

From Figure 9, we see that the P-A relations of the RRab stars in And XI and And XIII 
agree well with the general trend of the stars in the Brown field, while the RRc candidates 
in both dSph's are obviously shifted toward longer periods with respect to the M31 
RRc stars.
The mean RRab periods for And XI and And XIII are $<P_{ab}>$ =0.621 $\pm$ 0.026 (error1) 
$\pm$ 0.022 (error2) days and 0.648 $\pm$ 0.026 (error1) $\pm$ 0.022 (error2) days, 
respectively, while for the Brown field, this value becomes 0.594 days (B04). The ``error1'' value 
represents the standard error of the mean and the ``error2'' value is the error derived from our 
synthetic light curve simulations. The difference in the mean RRab periods between the M31 
halo and these two dSph's is mainly due to the lack of short period RRab stars with high 
amplitudes in And XI and And XIII.

To better understand the characteristics of the RRL populations of these two dSph's, we 
compare the P-A relations of And XI and And XIII with those of four other Andromeda dSph's 
(And I \& III (P05); And II (P04); And VI (P02)) in Figure 10. 
According to this comparison, the P-A relations of the RRL stars in And XI and And XIII show 
the closest resemblance to that of And III among these four Andromeda dSph's. 
The RRab stars in And III track the general trend of the Brown field RRab stars (B04). 
The lack of short period RRab stars with high amplitudes and the locations of the RRc stars 
in the P-A relation of And III mirror the characteristics of the RRL populations in And XI and 
And XIII. It is interesting to see that And III is the least luminous galaxy among these four  
dSph's and its mean metallicity ($<[Fe/H]>$=--1.7) is comparable to those of And XI and 
And XIII ($<[Fe/H]>_{And XI}$=--1.74; $<[Fe/H]>_{And XIII}$=--1.75, see section 5).  
It seems reasonable to assert that the behavior of the P-A relations of these three 
dSphs (And III, And XI, and 
And XIII) and their similar global properties could be related.
From Figure 10, we also see that the P-A relations of And I, II, III, and VI are as diverse 
as their complex star formation histories (Mateo 1998). The broad dispersion in the 
RRab periods of each Andromeda dSph also reflects the significant metallicity dispersions 
in these galaxies supporting previous assertions that these Andromeda dSph galaxies 
have likely experienced extended and complex chemical enrichment histories during 
the course of their evolution (P02, P04, P05).

\begin{figure}
\includegraphics[width=8cm]{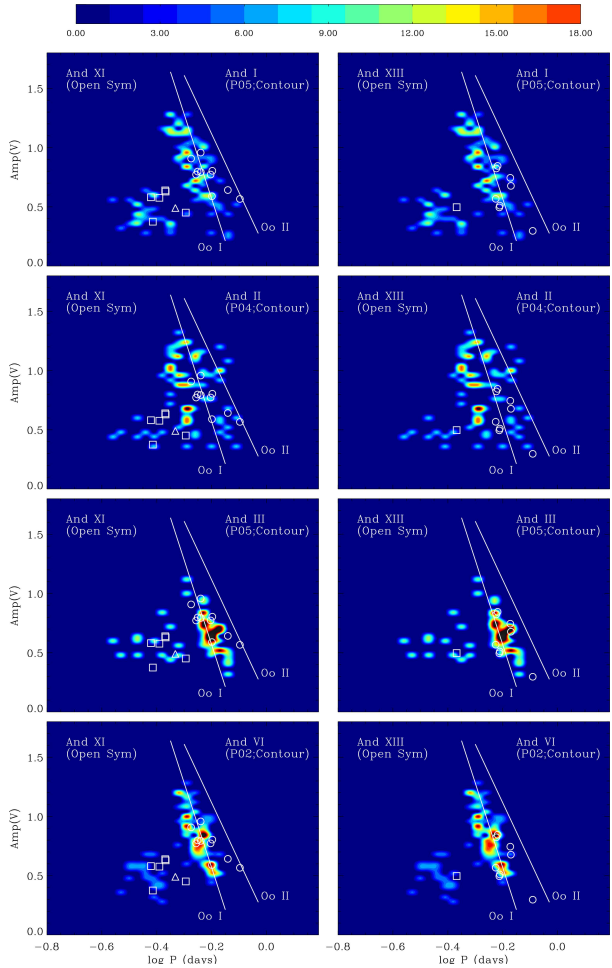}
\caption{A comparison of the P-A diagrams for the M31 dSph galaxies. See text for discussion.\label{fig10}}
\end{figure}

\section{Metallicity}
M06 derived the mean metallicities of And XI and And XIII by
interpolating between Padova isochrones (Girardi et al. 2004) plotted on the CFHT/MagaCam 
survey data. The estimated mean metallicity values are $<[Fe/H]>$=--1.3 and --1.4 dex for 
And XI and And XIII respectively with an error of $\pm$0.5 dex. Most recently, from their 
Keck/DEIMOS and SUBARU/SuprimeCam study of the faint M31 satellites, C10 estimated 
metallicities of individual RGB stars in these two dSph's via isochrone interpolation using 
the Dartmouth isochrones (Dotter et al. 2008) as well as Calcium II triplet spectra. The 
mean metallicities obtained by applying these two different methods converged at 
$<[Fe/H]>$=--2.0 $\pm$ 0.2 dex for both And XI and And XIII. C10 explained that the 
relatively higher mean metallicity values for these dSph's in the previous work of 
M06 were probably caused by the use of the Padova models, which systematically 
produce more metal-rich values in the MegaCam filters than other isochrone models.

We estimated the metallicities of And XI and And XIII via the period-amplitude-metallcity 
relationship for RRab stars derived by Alcock et al. (2000). The relation 
is given by the following equation : 

\begin{displaymath}
\left[Fe/H\right] = -8.85 \left[\log P_{ab} + 0.15 Amp\left(V\right)\right] - 2.60.
\end{displaymath}

The estimated mean metallicity of the RRab stars in And XI is 
$<[Fe/H]>$ = --1.75 $\pm$ 0.12 (error1) $\pm$ 0.13 (error2). For And XIII this value becomes 
$<[Fe/H]>$ = --1.74 $\pm$ 0.12 (error1) $\pm$ 0.14 (error2). Error1 represents the standard 
error of the mean, while error2 is the amount of error propagated from the determination 
of individual periods and amplitudes of the RRL candidates. Thus, the total errors in our 
metallicity determinations as described by the quadrature sum of error1 and error2 are 
$\pm$ 0.18 dex for both And XI and And XIII. 
To double check the validity of our estimates, we independently calculated
the mean metallicity of these dSphs using the S index (Hartwick 1968; Saviane et al. 2000). This
represent the slope of the RGB originally defined by the line connecting two points of the 
RGB in the (V,B--V) plane - the first point is at the level of the HB, and the other one is at 2.5 mag 
brighter than the HB. Saviane et al. (2000) redefined this relationship in the (V, V--I) plane 
by adjusting the second point to 2.0 mag brighter than the HB. The metallicity-S index 
relationship calibrated by Saviane et al. (2000) is given in the following form :

\begin{displaymath}
\left[Fe/H\right] = -0.24\times S + 0.28 \pm 0.12 \left(rms\right)
\end{displaymath}

Figure 11 illustrates our measurements of the S indices for And XI and And XIII.  The
(V--I) colors of RGBs at V(HB) and V(HB--2.0) were calculated from a least squares 
fit of a quadratic polynomial to the RGBs of each galaxy. By using the metallicity-S index
relationship above, we obtained [Fe/H] = --1.70 and --1.63 dex for And XI and And XIII,
respectively. The metallicity estimates for both And XI and And XIII from the RGB slope 
show excellent agreement with the RRL metallicities. It is worth mentioning here that 
we are aware of a caveat about the use of the MACHO relationship (Alcock et al. 2000). 
According to Cacciari et al (2005), this method can underestimate the metallicity especially 
for RRLs that have significantly evolved from the zero-age horizontal branch. 
This evolutionary effect generally raises the luminosity of RRL stars. Then, the pulsation 
period of these evolved RRLs becomes longer, as a result, yielding a relatively lower metallicity. 
Our metallicity estimates based on these two completely independent methods agree
very well and differ by only $\sim$ 0.1 dex in the worst case. Therefore we conclude that the 
RRL evolutionary effect does not negatively impact our metallicity determination using the 
MACHO relationship for And XI and And XIII. 

\begin{figure}
\includegraphics[width=8cm]{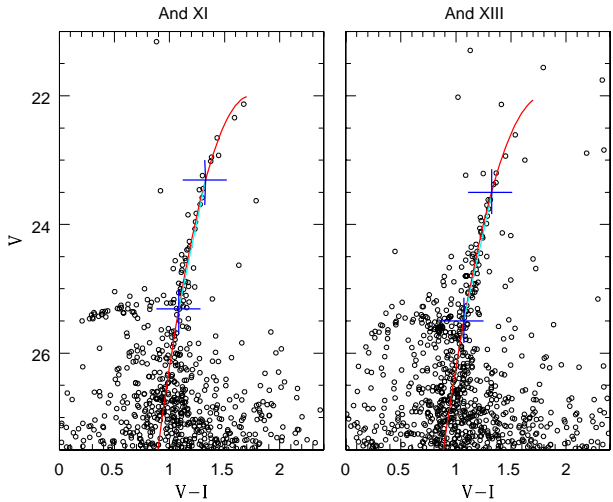}
\caption{The illustration of the S index method for the measurement of mean metallicity
(Saviane et al. 2000). 
The red solid lines represent RGB fiducials for each galaxy derived from a least square fit of 
a quadratic polynomial. The points marked as large plus symbols are the V magnitude of the
RGB at the HB level and that of the RGB at 2.0 mag brighter than the HB level.\label{fig11}}
\end{figure}

\section{Reddening and Distance}
Considering the brightest RGB star confirmed by their Keck/DEIMOS spectroscopy as a 
proxy for the tip of the RGB (TRGB) in each galaxy, C10 set distance ranges of 610--770 kpc, 
and 750--940 kpc for And XI and And XIII respectively.
For this calculation, they assumed that the brightest DEIMOS star is found within 
0.5 mag of the TRGB. C10 also provided distances for And XI and And XIII estimated 
by utilizing their HB magnitudes. In this case, the distances of And XI and And XIII become 
$\sim$ 830 kpc, and 910 kpc, respectively.


The previous studies (M06 and C10) used a reddening value of E(B-V)=0.08 taken
from the Galactic extinction map of Schlegel et al. (1998) in the calculation
of the distances for And XI and And XIII.
To double check whether the use of this reddening value is reasonable, we have
calculated the metallicity values for each dwarf galaxy by using this
reddening in the
relationship between metallicity and intrinsic RGB color at the level of HB
(i.e. $(V-I)_{0,g}$) derived by Saviane et al. (2000).

The observed RGB colors at the level of HB ($(V-I)_{RGB}$) for And XI and And XIII
are 1.108 $\pm$ 0.020 and 1.061$ \pm$ 0.013 (see Table 3 in the following section) respectively.
Then the adopted reddening value ($E(B-V)=0.08$) was converted into $E(V-I)$ using
$E(V-I)=1.38E(B-V)$ (Tamman et al. 2003) in order to calculate $(V-I)_{0,g}$.
The metallicity-$(V-I)_{0,g}$ relationship presented by Saviane et al. (2000) is
given in the following form :

\begin{displaymath}
\left[Fe/H\right] = 5.25\times(V-I)_{0,g} - 6.52.
\end{displaymath}

It should be noted that the metallicity used for this calibration is on the 
Zinn \& West scale (Zinn \& West 1984).
The adopted reddening yields the resultant metallicity
values of $[Fe/H]_{And XI} = -1.28 \pm 0.11$ and $[Fe/H]_{And XIII} = -1.53 \pm 0.07$ 
both of which 
are significantly more metal-rich than not only our estimates via the
RRL period-amplitude relation and the RGB slope, but also the metallicity values of 
M06 and C10, which used the reddening value of $E(B-V)$=0.08. Especially in
the case of And XI, these metallicities are clearly not consistent with
the overall properties seen in the CMDs and RRL populations of these galaxies. 
This result suggests the need for a more robust estimate of the reddening toward 
And XI and And XIII.

Using the metallicity values derived from the RGB slopes of And XI and XIII, we 
calculated $(V-I)_{0,g}$ using the metallicity-$(V-I)_{0,g}$
relation from Saviane et al. (2000). These $(V-I)_{0,g}$ values were then
combined with the observed $(V-I)_g$ quantities and
the conversion between $E(V-I)$ and $E(B-V)$ (Tamman et al. 2003, see above) 
to give reddening values of $E(B-V)_{And XI}$=0.15 
and $E(B-V)_{And XIII}$=0.11.



Moving on now to calculate the distances, we note that the mean V magnitude 
of 11 RRab candidates found in And XI is $<V(RR)>$ = 25.31 $\pm$ 0.02.    
By applying the extinction law, Av = 3.1E(B-V) = 0.450, we obtain a mean 
intrinsic V magnitude for the And XI RRab stars of $<V_{0}(RR)>$ = 24.86 $\pm$ 0.02. Then, the 
absolute V magnitude of each And XI RRab star is calculated using the metallicity-luminosity 
relationship for RRL stars, $M_{V}(RR)$ = 0.23 [Fe/H] + 0.93, from Chaboyer (1999). This yields a 
mean absolute V magnitude of$<M_{V}(RR)>$ = 0.53 $\pm$ 0.05. The quoted errors represent the 
quadrature sum of the standard error of the mean (0.03 mag) and the amount of error propagated 
from our metallicity determination (0.04 mag). 
Thus, we find the mean absolute distance modulus for And XI to be $(m-M)_{0}$ = 24.33 $\pm$ 0.05. 
This places And XI at a distance of $\sim$ 734 $\pm$ 18 kpc, which is consistent with the distance range of 
610--770 kpc obtained from the TRGB method, but smaller than the distance of 830 kpc 
estimated using the HB magnitude with smaller reddening value in the previous work of C10. 

For And XIII, the apparent mean V magnitude of 8 RRab candiates 
is $<V(RR)>$ = 25.49 $\pm$ 0.02. The estimated extinction, Av = 0.341 yields the intrinsic V magnitude
for the And XIII RRab stars of $<V_{0}(RR)>$ = 25.49 $\pm$ 0.02.
Employing once again the Chaboyer (1999) equation for the RRab metallicity-luminosity relation,  
we find $<M_{V}(RR)>$ = 0.53 $\pm$ 0.05. This gives a distance modulus of 
$(m-M)_{0}$ = 24.62 $\pm$ 0.05 (D $\sim$ 840 $\pm$ 18 kpc) for And XIII, placed in the middle
of the distance range of 750 -- 940 kpc estimated in the previous study of C10.




\section{Horizontal Branch Morphologies}

We make use of a newly quantified HB index, $\Delta$(V-I) (Dotter et al. 2010, hereafter D10), in order to 
characterize the HB morphologies of And XI and And XIII. This index is defined as the difference in 
color between the HB and the RGB at the level of the HB. The advantage of using this metric in the 
measurement of HB morphology is twofold. First, the $\Delta$(V-I) index has less degeneracy at 
both (red \& blue) ends of the HB morphology range (see Fig. 2 of D10) than the conventional 
HB index, (B-R)/(B+V+R) (Lee et al. 1994), which relies on counting stars on the blue (B) and 
red (R) portions of the HB as well as the RR Lyrae variables (V). Second, the calculation of  
$\Delta$(V-I) is relatively straightforward thereby minimizing the uncertainties associated with 
the census of RR Lyrae variables and the contamination of the red HB by intermediate-age RC stars. 

\begin{figure}
\includegraphics[width=8cm]{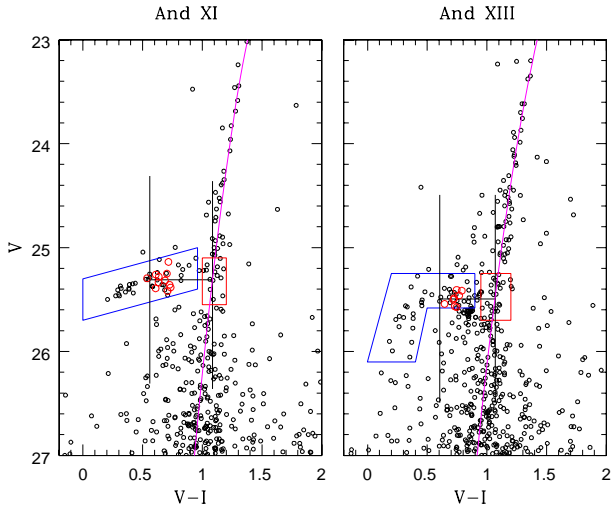}
\caption{These panels show the measurement of the $\Delta$(V-I) index. The blue and red boxed 
regions describe the selection of HB and RGB stars, respectively. The horizontal line presents the 
level of the horizontal branch obtained from the mean V magnitude of RRab stars. The vertical 
lines indicate the median colors of the HB and RGBs. It should be noted that RRab stars were not included in the calculation of the median color of the HB (see D10 for more detail). The dotted lines are RGB fiducials for each 
galaxy obtained by fitting a quadratic polynomial.
\label{fig12}}
\end{figure}

Figure 12 illustrates the measurement of the $\Delta$(V-I) indices for And XI and And XIII. The mean 
V magnitude of the HB, $<V(HB)>$, was obtained from the mean V magnitude of RRab stars, 
$<V(RR)>$,  calculated from the best fitting light curves ($<V(RR)>_{And XI}$ = 25.31; 
$<V(RR)>_{And XIII}$ = 25.49,  see section 4.1). Stars inside the blue-boxed regions were 
considered to be HB stars, while RGB stars at the level of the HB were selected within the 
red rectangles in the VI CMDs of each galaxy. The difference between the median colors 
of HB and RGB stars was calculated to yield $\Delta$(V-I) indices for each galaxy. The 
uncertainties in the $\Delta$(V-I) indices were calculated by the quadratic sum of the 
1-$\sigma$ errors in the median colors of the HB and RGB. We used a bootstrapping method 
for estimating errors in the median colors. Ten thousand  re-samples of the HB and RGB stars 
were constructed by using random sampling with replacements from the observed HB and RGB 
data. The standard deviation of the bootstrapped medians represents the 1-$\sigma$ errors in 
the median colors of the HB and RGB. The HB parameters for each galaxy are listed in Table 3.


\begin{figure}
\includegraphics[width=8cm]{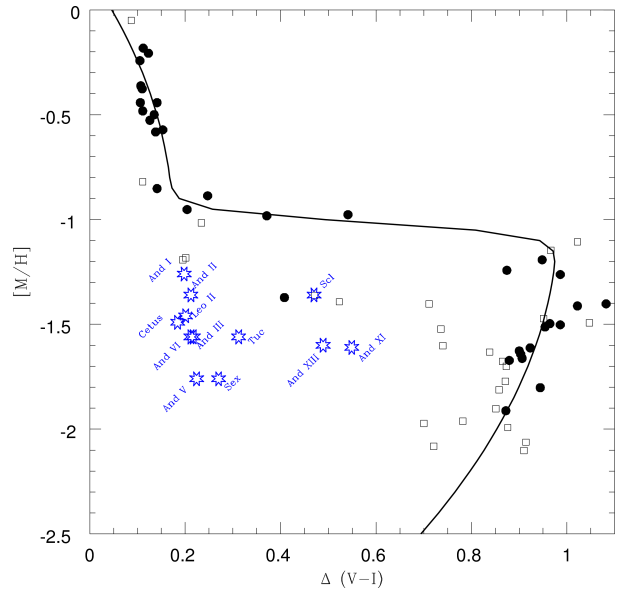}
\caption{The metallicity-HB morphology diagram for 65 Galactic Globular Clusters from D10 is 
presented. Solid circles and open squares represent inner ($R_{gc} < $ 8 kpc) and outer 
($R_{gc} > $ 8 kpc) halo GGCs, respectively. The solid line is the fiducial fit for the inner GGCs from D10. 
And XI and And XIII are marked as blue open stars along with 10 other Local Group dSph 
galaxies (Harbeck et al. 2001; Sarajedini et al. 2002; Komiyama et al. 2007). 
The Local Group dSph galaxies generally exhibit redder HBs than
the GGCs at the same metal abundance.
\label{fig13}}
\end{figure}

The [M/H] vs $\Delta$(V-I) diagram for 65 GGCs (D10) is shown in Figure 13. Solid circles and 
open squares represent inner ($R_{GC} < $8 kpc) and outer ($R_{GC} > $8 kpc) halo GCs, 
respectively. And XI and And XIII are marked as blue open stars along with 10 other Local Group 
dSph galaxies (Harbeck et al. 2001; Sarajedini et al. 2002; Komiyama et al. 2007). 
The solid line is an empirical fit to the inner halo GCs from D10. 
We adopted the [Fe/H] values of And XI and And XIII from section 5 of the present paper in which 
the $[Fe/H] - log P_{ab} - Amp(V)$ relationship from Alcock et al (2000) is used for the metallicity 
derivation. Then, these [Fe/H] values are translated into [M/H] by applying the following relation, 
$[M/H] = [Fe/H]$ $+$ $\log_{10}(0.638\times10^{[\alpha/Fe]}+0.362)$ from Salaris et al. 
(1993) by assuming $[\alpha/Fe]$=0.2 
(Tolstoy et al. 2003) for both galaxies. For the other 8 dSph galaxies, the metal abundance values are 
taken from the compilation of Grebel, Gallagher \& Harbeck  (2003); the corresponding [M/H] values 
are calculated with the same procedure used for And XI and XIII. Since the HB indices of the 
additional 8 dSph galaxies are in the canonical form of (B-R)/(B+V+R), these values need to be converted into the new $\Delta$(V-I) index. Figure 14 illustrates the correlation between (B-R)/(B+V+R) and $\Delta$(V-I) derived by a least squares fit of a cubic polynomial to the data set obtained from Figure 2 of D10. We have used this relation to calculate the $\Delta$(V-I) values for the 8 dSph galaxies. To double check the validity of this relationship, we measured (B-R)/(B+V+R) indices for And XI and And XIII (see Figure 15) and plotted them in the (B-R)/(B+V+R) vs $\Delta$(V-I) relation with an open star symbol. For the (B-R)/(B+V+R) index calculation, the boundaries of the instability strip in the VI CMDs were determined by the blue and red edges of the RRL (V-I) colors. From Figure 14 we see that the locations of And XI and And XIII agree well with the derived relationship (dashed line). We summarize the HB indices and other physical parameters of our sample dSph galaxies in Table 4.

\begin{figure}
\includegraphics[width=8cm]{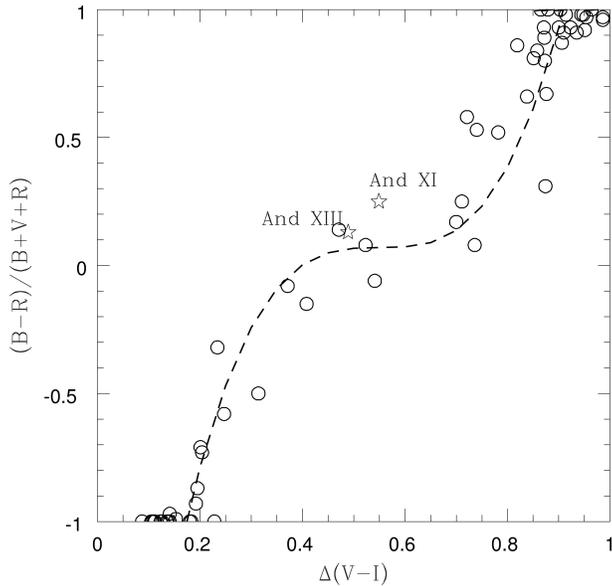}
\caption{The correlation between $\Delta$(V-I) and (B-R)/(B+V+R) index (see Figure 2 of D10). 
The dashed line illustrates the least square fit of a cubic polynomial to the data. We used this relation 
to convert the (B-R)/(B+V+R) values of 8 other Local Group dSph galaxies obtained from Harbeck et al. (2001).
\label{fig14}}
\end{figure}

\begin{figure}
\includegraphics[width=8cm]{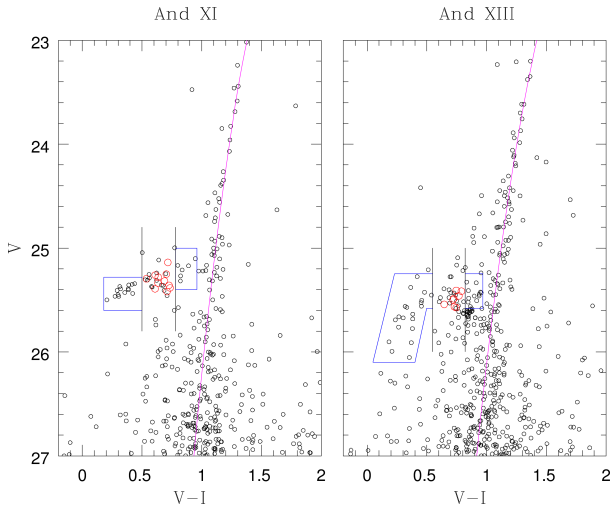}
\caption{The measurement of (B-R)/(B+V+R) indices for And XI and And XIII. The blue and red HB stars are selected by the enclosed regions. The boundaries of the instability strip were obtained from the (V-I) colors of RRL stars.
\label{fig15}}
\end{figure}

\begin{table}
\begin{center}
\caption{HB parameters \label{tbl-3}}
\begin{tabular}{ccccc}
\hline
\hline
Object   & $<V(HB)>$              & $(V-I)_{HB}$       & $(V-I)_{RGB}$      & $\Delta (V-I)$    \\
         &                            &  median            &   median           &                   \\
\hline
And XI   &  25.31$\pm$0.02            & 0.559$\pm$0.070  & 1.108$\pm$0.020  & 0.549$\pm$0.074 \\
And XIII &  25.49$\pm$0.02            & 0.572$\pm$0.109  & 1.061$\pm$0.013  & 0.489$\pm$0.110 \\
\hline
\end{tabular}
\end{center}
\end{table}

\begin{figure}
\includegraphics[width=8cm]{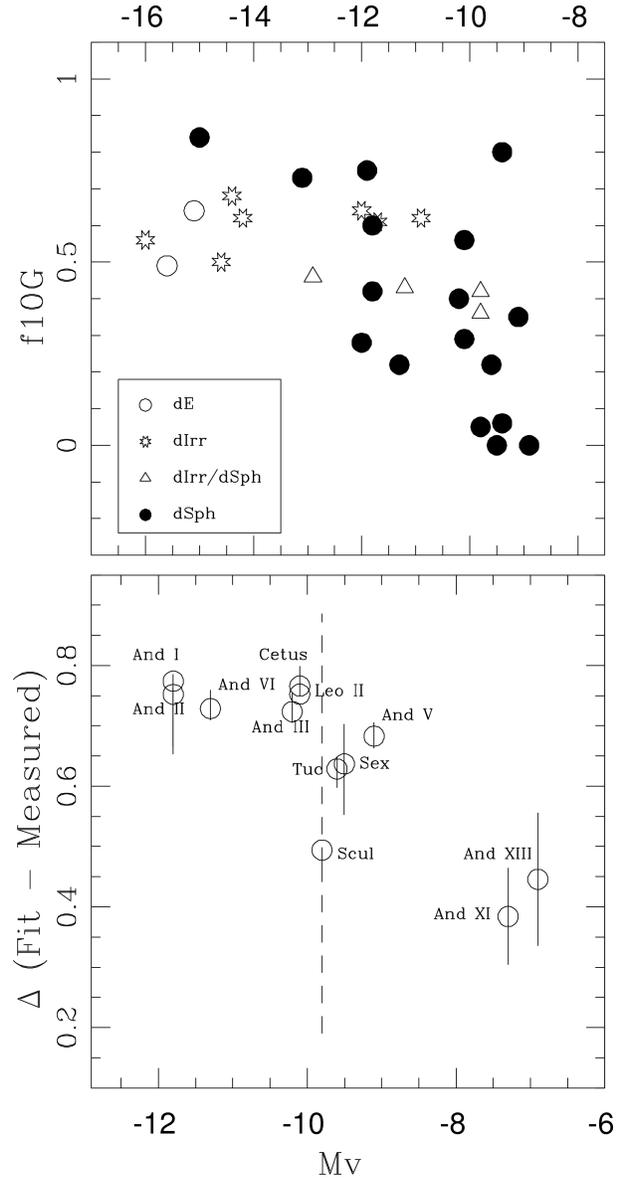}
\caption{The upper panel shows the relation between the star formation history of
dwarf galaxies over the past 10 Gyr ($f_{10G}$) from the work of Dolphin et al.
(2005) and their integrated absolute magnitudes. For the lower panel, we measure
the difference between each dwarf galaxy's HB index and that of the inner halo globular
clusters in Figure 13 [$\Delta$(Fit -- Measured)] and plot that as a function of the
galaxy's absolute magnitude.
\label{fig16}}
\end{figure}

From Figure 13, we see that the Local Group dSph galaxies and the GGCs show different behavior
in the metallicity-HB morphology diagram; that is to say the dSph galaxies exhibit
HBs that are generally redder than 
those of the GGCs at the same metallicity. This suggests that the stellar populations in these Local 
Group dwarf galaxies are subject to the second parameter effect. If we accept the conventional age 
hypothesis (Gratton et al. 2010; D10, and references therein), the metallicity-HB index diagram 
indicates that the stellar 
populations in these Local Group dwarf galaxies are $\sim$ 1-3 Gyr younger than the GGCs in 
terms of their mean ages (D10). However, unlike GGCs, the vast majority of which are single-aged systems,
the dSph galaxies are usually multiple-generation systems with various stellar populations formed 
through extended star formation and complex chemical enrichment processes. Considering this 
fundamental difference between the Local Group dwarf galaxies and GGCs, age itself is not likely to
be enough to fully explain the diversity in the HB morphologies of these dwarf 
galaxies. The metallicity-HB morphology relation of the Local Group dSph satellites likely 
reflects a combination of effects from multiple parameters such as [$\alpha/Fe$], helium 
content, mass loss, and environmental conditions (e.g. photo-ionizing radiation from host galaxies) -
in other words, their star formation histories. In fact, from Figure 13, it would appear that dSphs with a lack of recent star formation events tend to show bluer HB morphologies (Sculptor, And XI, and And XIII), 
while the dSphs with extended and more recent star formation exhibit generally redder HBs. 
For example, according to Dolphin et al. (2005), over 90\% of the stars in the Sculptor dSph formed 
more than 10 Gyr ago and subsequent star formation has largely ceased so that there are few signs
of recent star formation events. As a result, the HB morphology of the Sculptor dSph remains 
mostly blue and has not been significantly contaminated by intermediate aged RC stars compared 
to other dSphs. In addition, as mentioned in the previous section, both And XI and And XIII show no
indications of intermediate or young stellar populations in their CMDs. As such, their CMDs are reminiscent
of purely old, metal-poor GGCs. 

In order to probe this effect in more detail, we note that Dolphin et al. (2005) find
that for dSph and dwarf elliptical systems, the more intrinsically luminous galaxies 
generally exhibit extended star formation episodes, while the less luminous ones 
are more likely to have formed the bulk of their stellar populations over 10 Gyr ago. This
is graphically illustrated in the upper panel of Figure 16 where we have plotted the data
in Table 1 of Dolphin et al. (2005). The $f_{10G}$ quantity represents the fraction of
the stellar population in each galaxy that has formed between the present and
10 Gyr ago. As such, a low value of $f_{10G}$ means that there has been little
recent star formation. The upper panel of Figure 16 suggests therefore that there
is a correlation between absolute magnitude and the amount of recent star formation.

To examine the validity of our assertion - that the degree of recent star formation 
is responsible for the location of the dwarf galaxies to the left
of the inner halo GC locus in Figure 13, we plot the deviation of each galaxy from
this locus as a function of the galaxy's $M_V$ (which is a reflection of the degree of
recent star formation) in the lower panel of Figure 16. The error bars represent the metallicity
errors of each system propagated through to errors in $\Delta$(Fit -- Measured)
(cf. Dotter et al. 2010). The lower panel of Figure 16 reveals that
the deviation of each dSph from the inner halo line in Figure 13 seems to depend on its
$M_V$ (i.e. the degree of recent star formation). Thus, we have affirmed our assertion
that the deviation of the dSphs from the inner halo line is likely to be due to
their more complex star formation histories as compared to the majority of Galactic GCs.

The one possible outlier in the lower panel of Figure 16 is the Sculptor dSph, which seems 
to have had a fairly quiescent star formation history in contradistinction to that implied by its $M_V$.
This could imply that star formation in Sculptor was quenched by an external process
such as gas stripping via a dynamical interaction with the Milky Way or another more
massive galaxy.

\begin{table*}
\centering
\begin{minipage}{100mm}
\caption{HB index and other physical parameters of sample dSph galaxies. \label{tbl-4}}
\begin{tabular}{clccccccc}
\hline
               &
 Object        & (B-R)/(B+V+R) &
$\Delta (V-I)$ & [Fe/H]        &
               & [M/H]         &
               & $M_{V}$          \\
\hline
  & Sculptor &  0.06  &  0.470  &  -1.5 $\pm$ 0.5  &   &  -1.40  &  &  -9.8    \\
  & Sextans  & -0.37  &  0.270  &  -1.9 $\pm$ 0.4  &   &  -1.80  &  &  -9.5    \\
  & Tucana   & -0.20  &  0.312  &  -1.7 $\pm$ 0.2  &   &  -1.60  &  &  -9.6    \\
  & And I    & -0.80  &  0.198  &  -1.4 $\pm$ 0.2  &   &  -1.30  &  & -11.8    \\
  & And II   & -0.70  &  0.212  &  -1.5 $\pm$ 0.3  &   &  -1.40  &  & -11.8    \\
  & And III  & -0.67  &  0.217  &  -1.7 $\pm$ 0.2  &   &  -1.60  &  & -10.2    \\
  & And V    & -0.62  &  0.224  &  -1.9 $\pm$ 0.1  &   &  -1.80  &  &  -9.1    \\
  & And VI   & -0.70  &  0.212  &  -1.7 $\pm$ 0.2  &   &  -1.60  &  & -11.3    \\
\\
  & And XI   &  0.25  &  0.549  &  -1.75 $\pm$ 0.12 &  &  -1.65  &  & -7.3    \\
  & And XIII &  0.13  &  0.489  &  -1.74 $\pm$ 0.12 &  &  -1.64  &  & -6.9    \\
\hline
\end{tabular}
\end{minipage}
\end{table*}

\section{Discussion}

The luminosity-metallicity (L-M) relation of the Local Group dwarf satellites reflects the 
intrinsic properties of dwarf galaxies and contains significant implications for the 
mechanism of galaxy formation and the environmental conditions present during the 
early epochs of star formation (Harbeck et al 2005; Bovill \& Ricotti 2009).

The standard scenarios of galactic chemical enrichment (Larson 1974; Tinsley \& Larson 1979;
Dekel \& Silk 1986) suggest that a L-M relation should be present for dwarf galaxies.
The L-M relation shows that 
more luminous (i.e. more massive) galaxies are more metal-rich while less luminous 
(i.e. less massive) ones tend to be more metal-poor (Mateo 2008).
This agrees very well with the theoretical prediction that more massive systems largely retain
their interstellar media in the face of galactic winds triggered by supernova explosions
so that more metals are trapped in
these dwarf galaxies (Saviane et al. 2008). The L-M relation of the Local Group dwarf 
galaxies has also revealed the fundamental differences and possible correlations between 
dwarf irregular (dIrr) and dSph systems (Skillman \& Bender 1995; Richer et al. 1998; Mateo 1998; G03).
From their comparative study of 40 nearby dwarf galaxies,
G03 noted that the old stellar populations of dIrrs are systematically more metal-poor as 
compared with those of dSphs at the same luminosity
(i.e. present-day dIrrs are at least 10 times brighter in luminosity than
present-day dSphs at a given metallicity). This indicates that the gas deficient dSphs
in which recent star formation has been largely dormant, should be more effectively
enriched than dIrrs, which currently exhibit on-going star formation.
Based on the present-day luminosity, H I gas content, and the modest rates of on-going
star formation of the dIrrs, G03 concluded that most of the present-day
dIrrs in the Local Group, cannot fade enough to evolve into dSphs over a Hubble
time. Hence, the dIrrs are not likely precursors of the dSphs; furthermore, these two classes
of galaxies are fundamentally different stellar systems that have experienced different
paths in their evolution.

\begin{figure}
\includegraphics[width=8cm]{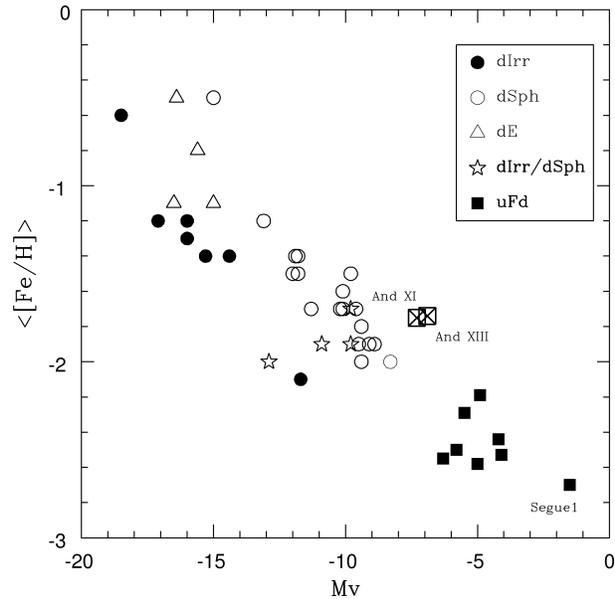}
\caption{Luminosity-metallicity (L-M) relation for Local Group dwarf galaxies.
The data were obtained from the compilation of Grebel, Gallagher, \& Harbeck (2003) for the canonical 
dwarf satellites ($M_{V} < -8 $). For the uFd systems, the data come from the recent work of 
Norris et al. (2010). And XI and And XIII are shown as marked boxes. The metallicity-$M_{V}$ relation 
appears to be very well established from the brightest dE, M32, to the faintest MW satellite, Segue 1. 
There is a noticeable gap around $M_{V}$ = --6 $\sim$ --8. And XI and And XIII seem to fill this gap 
but they exhibit relatively higher metallicities than the mean L-M relation at a given absolute V
magnitude. The typical error in the mean metallicity is about $\sim$ 0.3 -- 0.4 dex.
\label{fig17}}
\end{figure}

With our newly determined metallicities from the RRL periods, we have plotted And XI and And XIII 
on the L-M relation of the Local Group dwarf galaxies in Figure 17. The absolute V magnitudes
of And XI and And XIII were adopted from the work of C10.
From among the 40 nearby galaxies in the compilation of G03, we have included 34 Local Group
dwarf satellites within 1.5 Mpc of the Milky Way (MW). The L-M relation for 8 recently discovered
ultra-faint dwarf (uFd) satellites of the MW were obtained from the recent work of Norris et al. (2010). 
The ordinate of this plot represents the mean metallicity of the old stellar populations in 
each galaxy while the abscissa is the integrated absolute V magnitude of each dwarf 
galaxy. To better understand this plot, a different symbol was used for each class of dwarf galaxy 
(solid circles (dIrr); open circles (dSph); open triangles (dE); open stars (dIrr/dSph,
transition type); solid squares (ultra faint dwarf)). From Figure 17, we see that even given 
a typical metallicity error of $\sigma_{<[Fe/H]>}$ $\sim$ $\pm$0.4 dex, a well established
L-M relationship exists from the brightest dIrr and dE to the faintest dwarf satellite,
Segue 1. However, there is a noticable gap between the faint end of the canonical dwarf galaxies
and the bright end of the newly discovered ultra-faint dwarf satellites.
We also observe the well known features in the L-M relationship that dSphs are more metal-rich
than the dIrrs at the same absolute V magnitude, while the transition type dwarf galaxies
(dIrr/dSph : Phoenix, DDO 210, LGS 3, and Antlia) show similar characteristics to the dSphs
(Skillman \& Bender 1995; Richer et al. 1998; Mateo 1998, and G03).
Our estimates of the metallicity ($<[Fe/H]>_{And XI}$=--1.75; $<[Fe/H]>_{And XIII}$=--1.74) 
tend to locate And XI and And XIII slightly off from the average L-M relation as compared with 
the previous work of C10 ($<[Fe/H>$=-2.0 for both And XI and And XIII). However
both galaxies still appear to follow the general trend of this L-M relation within the errors
of the metallicity determinations.

The observed properties of the newly discovered uFd galaxies, such as size, surface brightness,
mass to light ratio, and L-M relation can be explained within the context of  
``the tidal disruption scenario'' of dSph formation which suggests that the present-day faint 
galaxies were once
much more massive but lost significant mass via tidal interactions with their giant host galaxies
(e.g. Ursa Minor dSph : Martinez-Delgado et al. 2001; Mayer et al. 2001).
However, this tidal scenario cannot account for all of the properties of uFd galaxies, especially
the very metal-poor nature of some of the uFd galaxies. If the dSphs formed in massive halos,
the deep gravitational potential well of these halos should have effectively retained their metals. 
Hence, there should be a lower limit to the metallicity of the dSph galaxies.
The continuously decreasing metallicity in the L-M relation from the bright dIrrs and/or dEs 
to the uFd satellites may imply that the uFd galaxies were originally less massive stellar 
systems formed in low mass dark matter halos with fewer metals in which most of stars formed 
before the end of reionization ($6 < z < 20$; Bovill \& Ricotti 2009; Walker et al. 2009; 
Salvadori \& Ferrara 2009). In terms of the absolute V magnitude, And XI ($M_{V}$=--6.9) and 
And XIII ($M_{V}$=--6.7) appear to reside in the uFd regime (C10). As mentioned above, 
their metallicities seem to
be relatively higher than the average L-M relation at a given luminosity. This opens up the 
possibility that these two faint dSph may be tidal remnants from interactions with their giant 
host, M31 or prereionization fossils in which metallicity was enriched by a late phase gas 
accretion and star formation event well after reionization at a redshift z$ < $1--2 (Ricotti 2009).
However, considering the likelihood that the absolute V magnitudes of And XI and And XIII are 
lower limits (M06), we cannot ignore the possibility that these 
faint dSph could be canonical dSph galaxies. Therefore, the verdict about the true origin of 
And XI and And XIII is still tentative. 

\section{Summary and Conclusions}

We present a comprehensive study of the stellar populations in two faint M31 dwarf 
satellites, And XI and And XIII, using deep archival images from HST/WFPC2. 
Based on the analysis of the HB morphology and the properties of the RRL stars, 
we obtain the following results : 

1. Based on the appearance of their CMDs, And XI and And XIII appear to be purely 
old and metal-poor stellar systems and appear to have experienced relatively simple 
star formation histories as compared with their more luminous counterparts in the Local Group.
 
2. We have discovered and characterized RRL populations in both And XI 
($N_{ab}=10, N_{c}=5$) and And XIII ($N_{ab}=8, N_{c}=1$) using our new template 
light curve fitting routine (RRFIT). The Period-Amplitude relations of the RRL stars in these 
two M31 dwarfs appear to follow the P-A trend of the RRL population found in the 
Brown field (B04). The lack of AC populations reinforces our assertion that both And XI and And XIII are purely old stellar systems

3. The metallicities of the RRab stars were calculated via the P-A-[Fe/H] relationship of 
Alcock et al. (2000). The metallicities obtained in this way ($[Fe/H]_{And XI}=-1.75$; 
$[Fe/H]_{And XIII}=-1.74$) are consistent with the values calculated from the RGB slopes
indicating that our measurements are not significantly affected by the evolution of the
RRLs away from the zero age horizontal branch.

4. The distance to each galaxy was determined using the absolute V magnitudes of the
RRab stars. We find $(m-M)_{0,V}$=24.33$\pm$0.05 for And XI and 
$(m-M)_{0,V}$=24.62$\pm$0.05 for And XIII.  Our results agree very well with the measurements 
from previous studies (M06, C10). 

5. The HB morphologies of And XI and And XIII were evaluated with a new metric called the $\Delta$(V-I) index (D10); this indicates that both of these faint galaxies are purely old systems. Our comparative analysis of the HB morphology also revealed that the fraction of the ancient stellar populations ($>$ 10 Gyr) in a Local Group dSph galaxy can be estimated by its HB morpholgy index. 

6. The locations of And XI and And XIII in the luminosity-metallicity plane
follow the general trend for 
the Local Group dwarf galaxies and appear to fill the gap between the faint end of the 
canonical dSphs and the bright end of the newly discovered uFd systems. 
The metallicities of And XI and And XIII seem to be relatively higher than the average 
metallicity of uFd systems at the same luminosity. This may be an indication that these
dwarfs were more massive in the past having experienced significant mass loss via tidal 
interactions with their giant host M31.  Another possibility is that these systems are 
reionization fossils in which metallicity was enriched due to late phase gas accretion and 
star formation well after reionization (z $\sim$ 1--2). A third possibility is simply that
there are significant uncertainties in the integrated absolute V magnitudes of And XI 
and And XIII and that the true values are brighter than the measured ones.


\section*{Acknowledgments}

We acknowledge support from NASA through grant AR-12153.01-A from the Space 
Telescope Science Institute, which is operated by the Association of Universities for 
Research in Astronomy, Inc., for NASA under contract NAS5-26555.

\appendix

\section{The Special Case of And II}
In examining the appearance of Fig. 14, it seems that And II is the only M31 dwarf that exhibits 
a significant period spread among its ab-type RRLs in the P-A diagram. This was noted by 
P04 as well who suggested a correlation between the ab-type RRL period spread and the 
significant metallicity dispersion seen among the RGB stars in the And II CMD. We would like
to investigate the bimodality of this period distribution and to correlate the inferred metallicity 
of the two RRL populations to see if
it is consistent with the appearance and location of the And II RGB.

Figure A1 shows the P-A diagram for And II wherein we have fitted a straight line to the 
ab-type RRL relation. For each such star, we then calculate the offset between the 
measured period and the predicted period ($\Delta$Log P) from this least squares fit, 
which is plotted as a solid line in Fig. A1. The inset to Figure A1 shows the histogram of 
$\Delta$Log P values generated in this way.
Fitting a single Gaussian to this histogram (dashed line in Figure A1 inset) produces a reduced-$\chi^2$
value of $\sim$14 which suggests that there is a less than 1\% chance that the distribution of
$\Delta$Log P is unimodal. In contrast, fitting the $\Delta$Log P histogram with a bimodal
Gaussian yields a reduced-$\chi^2$ of $\sim$0.3, suggesting good agreement between
the two distributions. The peaks of the two Gaussians differ by 0.100 $\pm$ 0.006 dex in Log P,
which corresponds to a metallicity difference of $\sim$0.90 $\pm$ 0.05 dex in [Fe/H] according to the
Alcock et al. (2000) equation quoted above. 

\begin{figure}
\includegraphics[width=8cm]{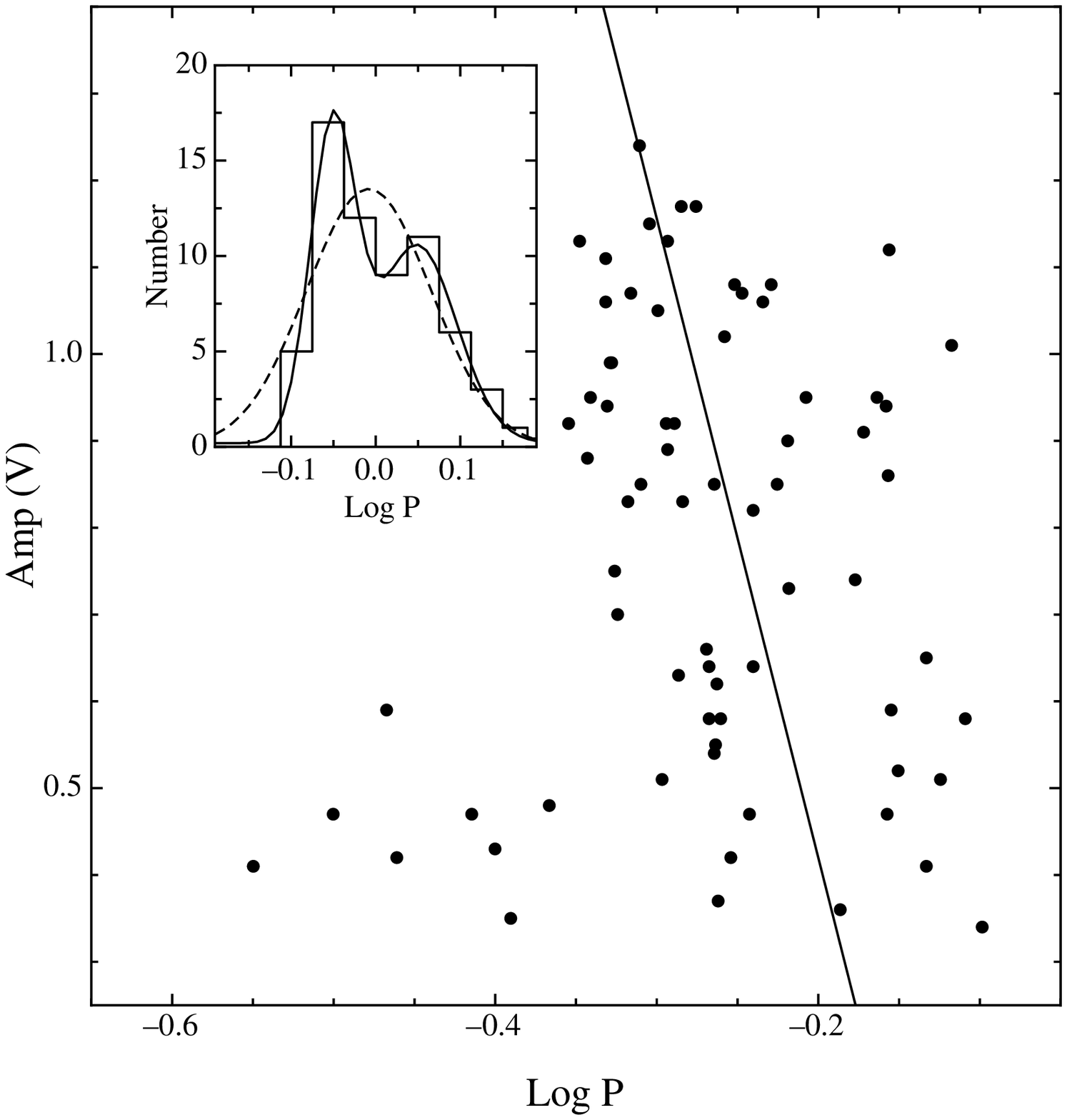}
\caption{The Period-Amplitude diagram for And II. The solid line represents a linear least 
square fit to the ab-type RRL relation. The inset histogram illustrates the distribution of 
$\Delta$ log P values which is the offset of the measured period for each RRab star 
from the least square fit. A single Gaussian fit to this histogram is shown as a dashed line, 
while a bimodal Gaussian is described by a solid line.\label{figa1}}
\end{figure}

Turning now to the appearance of the And II RGB, in Figure A2 we show the distance and reddening
corrected CMD of And II along with the standard globular cluster RGBs from Sarajedini \&
Layden (1997). The And II photometry represents our reduction of the program frames taken
for GO-6514 using the same procedure and techniques as described in Sec. 2 above.
The sequences for M15, NGC 6752, NGC 1851, and 47 Tuc are plotted as grey lines. According to
Zinn \& West (1984), these have metallicities of --2.17, --1.54, --1.29, and --0.71, respectively.
Our aim is to compare the metallicities from the RGB with the RRL values from the Alcock et al. (2000)
relation, which is also based on the Zinn \& West (1984) abundance scale. The GGCs have
been corrected for distance using the V(HB) and E(B--V) values from Sarajedini \&
Layden (1997) along with the RRL luminosity-metallicity relation from Chaboyer (1993, see below).
For And II, we adopt V(HB) = 24.93 and E(B--V) = 0.06 from values from Da Costa et al. (2000).

The appearance of the And II CMD suggests the presents of a bifurcation in its RGB brightward
of M$_V$ $\sim$ --1.5 with a break essentially midway between the NGC 6752 and NGC 1851 RGBs. 
As such, we have interpolated between these two RGBs and constructed a mean RGB represented by
the solid black line in Fig. A2. The solid line in the inset to Fig. A2 shows the histogram of color 
differences between this mean RGB sequence and each of the 134 RGB stars brighter than M$_V$=--1.5.
The dashed histogram in the inset is constructed using the original photometric data consisting of
74 RGB stars from Da Costa et al. (2000) showing good agreement with our result. A bimodal
Gaussian fit to the solid line in the Fig. A2 inset suggests a difference of 
$\delta$[$\Delta$(B--V)] = 0.352 $\pm$0.016 in the peaks of the two distributions. Using Equation (4)
of Sarajedini \& Layden (1997), this corresponds to a metallicity difference of $\sim$0.70 $\pm$ 0.03
dex in [Fe/H]. In terms of the random errors, this value is 3.4$\sigma$ different than the 
metallicity difference inferred from the ab-type RRLs above, but given the fact that the
systematic errors are likely to be larger than the random ones, the metallicity difference
inferred from the RGB stars and the ab-type RRLs are consistent with each other.

\begin{figure}
\includegraphics[width=8cm]{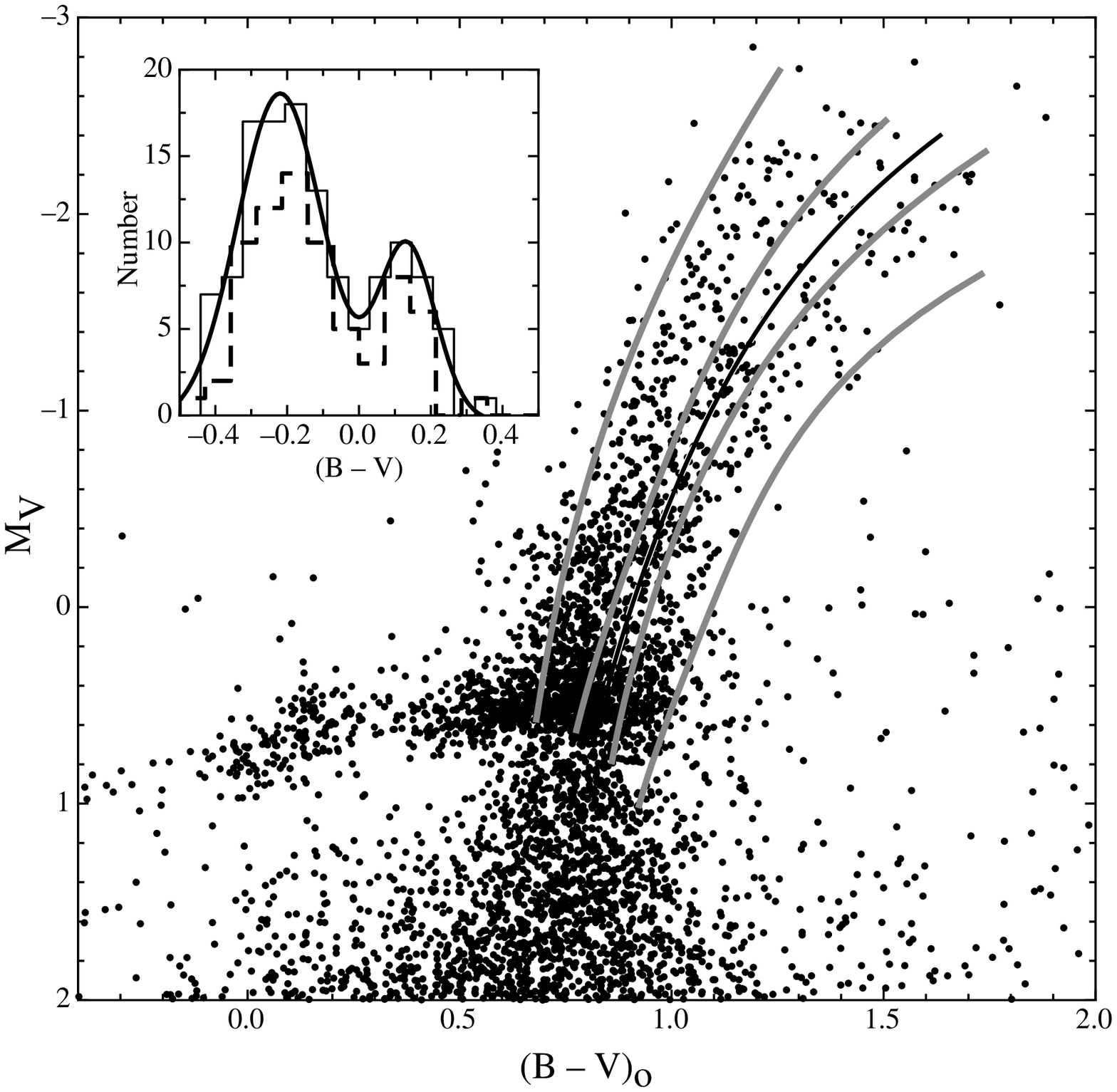}
\caption{The distance and reddening corrected CMD of And II. The RGB sequences for M15, 
NGC6752, NGC1851, and 47 Tuc obtained from Sarajedini \& Layden (1997) are illustrated as 
grey lines. The solid black line represents mid-points between the NGC6752 and NGC 1851 
RGBs, which is coincident with the bifurcation in the RGB of And II brightward of 
M$_{V}$ $\sim$ --1.5. The solid histogram in the inset figure illustrates the distribution of (B-V) color 
differences between this mean RGB sequence (solid black line) and And II RGB stars brighter than 
M$_{V}$=--1.5. The dashed histogram in the inset is constructed using the original And II data as
published by Da Costa et al. (2000). The solid line in the inset represents a bimodal Gaussian 
fit  to the solid histogram. \label{figa2}}
\end{figure}


\bsp

\label{lastpage}

\end{document}